\documentclass[onecolumn,showpacs]{revtex4}
\usepackage{graphicx}
\usepackage{dcolumn}
\usepackage{bm}
\input epsf
\begin{document}
\title{\Large  GENERALIZED SECOND LAW OF THERMODYNAMICS ON THE EVENT HORIZON FOR INTERACTING DARK ENERGY}
\author{\bf~Nairwita~Mazumder\footnote{nairwita15@gmail.com}, Subenoy~Chakraborty\footnote{schakraborty@math.jdvu.ac.in}.}
\affiliation{$^1$Department of Mathematics,~Jadavpur
University,~Kolkata-32, India.}
\date{\today}
\begin{abstract}
Here we are trying to find the conditions for the validity of the
generalized second law of thermodynamics (GSLT) assuming the first
law of thermodynamics on the event horizon in both cases when the
FRW universe is filled with interacting two fluid system- one in
the form of cold dark matter and the other is either holographic
dark energy or new age graphic dark energy. Using the recent
observational data we have found that GSLT holds both in
quintessence era as well as in phantom era for new age graphic
model while for holographic dark energy GSLT is valid only in
phantom era.
\end{abstract}
\maketitle
\section{\normalsize\bf{Introduction}}
The equivalence between a black body and a black hole(BH)both
emitting thermal radiation (using semi-classical description) put
a new era for black hole physics. The black hole behaves as a
thermodynamical system with the temperature (known as Hawking
temperature)and the entropy proportional to the surface gravity at
the horizon and the area of the horizon [1,2] respectively.
Further, this temperature, entropy and mass of a black hole are
related by the first law of thermodynamics [3]. On the other hand,
the thermodynamic parameters namely, the temperature and entropy
are characterized by the space time geometry. So a natural
speculation about some relationship between black hole
thermodynamics and Einstein equations is legitimate. In fact,
Jacobson [4] showed that Einstein equations can be derived from
the first law of thermodynamics :$\delta Q= TdS$ , for all local
Rindler Causal horizons with $\delta Q$ and $T$ as the energy flux
and Unruh temperature measured by an accelerated observer just
inside the horizon, while on the other way Padmanabhan [5] derived
the first law of thermodynamics on the horizon, starting from
Einstein equations for general static spherically symmetric
space-time.\\

This equivalence between the thermodynamical laws and the Einstein
gravity subsequently leads to generalize this idea in cosmology,
treating universe as a thermodynamical system. More precisely , if
we assume that the universe is bounded by the apparent horizon
$R_{A}$ with temperature $T_{A}=\frac {1} {2 \pi R_{A}}$ and
entropy $S_{A}=\frac{\pi {R_{A}}^2}{G}$ then the Friedmann
equations and the first law of thermodynamics (on the apparent
horizon ) are equivalent [6]. Usually, the universe bounded by the
apparent horizon is termed as a Bekenstein system because
Bekenstein's entropy-mass bound $(S\leq 2 \pi E R_{A})$ and
entropy-area bound ($S\leq \frac{A}{4}$) are obeyed in this region
. On the otherhand, the cosmological event horizon does not exist
in the usual standard big bang model while it exists in an
accelerating universe dominated by dark energy $\omega_{D}\neq -1$
. However , both the first and second law of thermodynamics break
down on the event horizon [7]. Using the usual definition of
temperature (as on the apparent horizon) Wang et al [7] have
argued that the applicability of the first law of thermodynamics
is restricted to nearby states of local thermodynamic equilibrium
while event horizon characterizes the global features of space
time.\\

The present observational evidences obtained from
Wilkinson-Microwave-Anisotropy-Probe(WMAP) strongly suggest that
the current expansion of the universe is accelerating [8,9]. There
are two possible ways [10-15] of explaining this accelerated
expansion of the universe.In the frame work of general relativity
it can be explained by introducing the dark energy having negative
pressure. The other possibility is to consider modified gravity
theory such as $f(R)$ gravity [15,16,17], where the action is an
arbitrary function $(f(R))$ of the scalar curvature $R$. As a
result, the Friedmann equations [18,19] become complicated by
including powers of Ricci scalar $R$ and its time derivatives. In
the present work, we examine the validity of the generalized
second law of thermodynamics of the universe bounded by the event
horizon (which exists due to present accelerating phase of the
universe). The matter in the universe is taken in the form of
interacting two fluid system- one component is dust and the other
is in the form of dark energy.\\

The model of dark energy obeying holographic principle is termed
as holographic dark energy. From the effective quantum field
theory, the energy density for the holographic dark energy is
given by [20] $$\rho_{D}=3c^2 M_{p}^{2} L^{-2},$$ where $L$ is an
IR cut-off in units $M_{p}^{2}=1$, c is any free dimension less
parameter,determined from observational data [21]. Li [22] has
argued that (also from the present context) the cut off length $L$
can be chosen as the radius of the event horizon to get correct
equation of state and desired accelerating universe. Very
recently, Wei and Cai [23] proposed another model of dark energy
known as new age graphic dark energy (NADE) which may have
interesting cosmological consequences (Originally Cai[24] proposed
an age graphic dark energy model(ADE) which is unable to describe
the matter dominated era). These new dark energy models are based
on the uncertainty relation in quantum mechanics and gravitational
effect due to Einstein gravity. Also the NADE models are
constrained by various astronomical observations [25].Although the
evolution behavior of the NADE has similarity [26] with the
holographic dark energy but the causality problem in the
holographic dark energy model can be overcome in ADE by choosing
the age of the universe as the measure of the length (instead of
the horizon distance) while the conformal time $'\eta'$ is chosen
as the time scale in NADE. Thus the energy density of the NADE can
be written as
\begin{equation}
\rho_{ND}=\frac{3n^2 m_p^2}{\eta^2}
\end{equation}
 where the conformal time $\eta$ has the expression
$$\eta = \int\frac{dt}a =\int_{0}^a \frac{da}{Ha^2}$$
and the numerical factor $3n^2$ is taken care of uncertainties
in quantum theory and the effect of curved space time.\\
The paper is organized  as follows. In the section (II) the
holographic dark energy model has been used while recently
formulated new age graphic dark energy model is taken in section
(III). The conclusions are presented in section (IV).\\

\section{\normalsize\bf{Interacting Holographic dark energy model and generalized second law of thermodynamics:}}
In this section, we take the FRW universe bounded by the event
horizon and the matter in the universe is taken as the holographic
dark energy (HDE) interacting with dust. So the energy density of
the HDE model has the expression
\begin{equation}
\rho_{D}=3c^2 M_{p}^{2} R_E^{-2}
\end{equation}
The individual continuity equations for the HDE and dust are of
the form
\begin{equation}
\dot{\rho_{D}}+3H(1+\omega_{D}) \rho_{D}=-Q
\end{equation}
and
\begin{equation}
\dot{\rho_{m}}+3H\rho_{m}=Q
\end{equation}
where $Q=\Gamma \rho_D$ [27] is the interaction term and the decay
rate $\Gamma$ corresponds to conversion of dark energy to dust.
The above conservation equations can be written in non-interacting
form as [27]
\begin{equation}
\dot{\rho_{D}}+3H(1+\omega_{D}^{eff}) \rho_{D}=0
\end{equation}
and
\begin{equation}
\dot{\rho_{m}}+3H(1+\omega_{m}^{eff})\rho_{m}=0
\end{equation}
These equations show that the interacting matter system is
equivalent to non-interacting two fluid system with variable
equation of state
\begin{equation}
\omega_D^{eff}=\omega_D+\frac{\Gamma}{3H} ~~~and~~~~
\omega_m^{eff}=-\frac{\Gamma}{3Hu}~
\end{equation}
where $~u=\frac{\rho_m}{\rho_D}~$ is the ratio of two energy
densities. So combining (5) and (6) we get
\begin{equation}
\dot{\rho_{t}}+3H(\rho_{t}+p_t)=0
\end{equation}
where
\begin{equation}
\rho_t=\rho_D+\rho_m ~,~p_D=\rho_D \omega_D^{eff}~,~p_m=\rho_m
\omega_m^{eff}~and~ p_t=p_D+p_m
\end{equation}
For FRW Universe with line element
$$
ds^{2}= -dt^{2}+\frac{a^2(t)}{1-kr^2}dr^2 + a^2(t) r^2
d\Omega_{2}^{2}
$$
the Friedmann equations are
 $$
H^{2}+\frac{k}{a^{2}}=\frac{8 \pi G}3 \rho_{t}
$$
and
\begin{equation}
\dot{H}-\frac{k}{a^{2}}=~-4 \pi G (\rho_{t}+p_{D})
\end{equation}
with~$\rho_{t}=\rho_{m}+\rho_{D}$~. Now as usual the density
parameters are
$$\Omega_m=\frac{\rho_m}{\frac{3H^2}{8 \pi G}}~,~\Omega_D=\frac{\rho_D}{\frac{3H^2}{8 \pi G}}~,
\Omega_k=\frac{\frac{k}{a^2}}{\frac{3H^2}{8 \pi G}}~$$ and due to
the first Friedmann equation they are related by the relation
\begin{equation}
\Omega_D+ \Omega_m=1+\Omega_k
\end{equation}
Now using the Friedmann equations, the conservation relations and
the expression for the energy density of the holographic dark
energy (Eq.(2)) , the equation of state parameter $\omega_D$ for
the HDE can be obtained as [27]
\begin{equation}
\omega_D=-\frac{1}3-\frac{2\sqrt{\Omega_D-\Omega_k}}{3c}-\frac{b^2(1+\Omega_k)}{\Omega_D}
\end{equation}
where the decay rate is chosen as [27]
\begin{equation}
\Gamma=3b^2(1+u)H
\end{equation}
with $b^2$ as the coupling constant.\\
The deceleration parameter $q=-(1+\frac{\dot{H}}{H^2})$ can be
expressed in terms of density parameters (using Friedmann
equations and the conservation relations) as
\begin{equation}
q=-\frac{\Omega_D}2 - \frac{\Omega_D \sqrt{\Omega_D-\Omega_k}}{c}
+\frac{1}2(1-3b^2)(1+\Omega_k)
\end{equation}
Now using the expression for the energy density of holographic
dark energy (Eq.(2)) and the modified conservation equation (5)
the change in the radius of the event horizon is given by [28]
\begin{equation}
dR_{E}=\frac{3}2R_{E}H(1+\omega_{D}^{eff})dt
\end{equation}
Assuming the validity of the first law of thermodynamics on the
event horizon and using the expression for the amount of energy
crossing the event horizon in time dt [6,29,30] i.e.
\begin{equation}
-dE=4\pi{{R}_E}^{3}H(\rho_t+p_t)dt
\end{equation}
we obtain
\begin{equation}
 dS_{E}=\frac{4\pi{{R}_E}^{3}H(\rho_t+p_t)dt}{T_{E}}
 \end{equation}
 where $S_{E}$ is the entropy of the event horizon and $T_{E}$ is the temperature on the event horizon.\\
 From the Gibb's equation [31]
 \begin{equation}
T_{E}dS_{I}=dE_{I}+p_tdV~,
\end{equation}
the variation of the entropy $(S_I)$ of the fluid inside the event
horizon is given by
\begin{equation}
\frac{dS_{I}}{dt}=\frac{4\pi
{R_{E}}^3}{T_{E}}H(\rho_{t}+p_{D})\left(\frac{3}2(\omega_{D}^{eff}+1)-1\right)
\end{equation}
In deriving Eq.(19) we have used Eq.(9) and the following
expressions $$V=\frac{4}3 \pi R_E^3~,~E_I=V \rho_t.$$
Hence combining equations (10) and (19) the resulting change of
total entropy is given by
$$
\frac{d}{dt}(S_{I}+S_{E})=\frac{6\pi {R_{E}}^3H}{T_{E}}
(\rho_{t}+p_{D})(\omega_{D}^{eff}+1)
$$
$$or~~~ more~~~ explicitly$$
\begin{equation}
=\frac{6\pi {R_{E}}^3H}{T_{E}}\left[ \rho_D
{(\omega_{D}^{eff}+1)}^2 + \rho_m
(\omega_{D}^{eff}+1)(\omega_{m}^{eff}+1)\right]
\end{equation}

\section{\normalsize\bf{Interacting New age graphic dark energy model and generalized second law of thermodynamics:}}
Similar to the previous section, the matter in the universe
bounded by the event horizon is taken as interacting two fluid
system- one component is in the form of recently formulated new
age graphic dark energy and the other is the dark matter in the
form of dust. So as before the time variation of the entropy of
the horizon can be obtained from the first law of thermodynamics
with expression
\begin{equation}
 dS_{E}=\frac{4\pi{{R}_E}^{3}H(\rho_t+p_t)dt}{T_{E}}
 \end{equation}
 where $\rho_t=\rho_{ND}+\rho_m ~,~p_{ND}=\rho_{ND}
\omega_{ND}^{eff}~,~p_m=\rho_m \omega_m^{eff}~and~
p_t=p_{ND}+p_m.$ Here the energy density of the dust component
$(\rho_m)$ satisfies the conservation equation (4)(or (6)) while
the NADE has energy density and pressure $\rho_{ND}$ and $P_{ND}$
with equation of state $P_{ND}=\rho_{ND} \omega_{ND}$. This matter
component satisfies the continuity equation (3) or (5). Also the
effective state parameters have the same form as in equation(7).
Further, an explicit form of $\rho_{ND}$ is given in equation (1)
in terms of conformal time. In contrast to HDE, the NADE energy
density (given in equation(1)) is not related to the radius of the
event horizon. So from the definition of the event horizon, the
time variation of $R_E$ is given by [32]
\begin{equation}
\frac{d{R}_E}{dt}=\left({R}_{E}-\frac{1}H\right)H
\end{equation}

Then using Gibbs equation, the time variation of the entropy of
the matter inside event horizon is given by
\begin{equation}
\frac{dS_{I}}{dt}=-\frac{4\pi {R_{E}}^2}{T_{E}}(\rho_{t}+p_{D})
\end{equation}
 Thus combining equation (21) and (23) the change of total entropy
 is given by

\begin{equation}
\frac{d}{dt}(S_{I}+S_{E})=4\pi
(\rho_t+p_t)\frac{{{R}_E}^{2}H}{T_{E}}\left({R}_{E}-\frac{1}H\right)
\end{equation}
 Further for the new age graphic dark energy the expressions of
 the equation of state parameters $\omega_{ND}$ and the
 deceleration parameter $'q'$ in terms of the density parameters
 are the following:
\begin{equation}
\omega_{ND}=-1+\frac{2\sqrt{\Omega_{ND}}}{3na}-\frac{b^2(1+\Omega_k)}{\Omega_{ND}}
\end{equation}
and
\begin{equation}
q=-\frac{3 \Omega_{ND}}2 - \frac{{\Omega_{ND}}^{\frac{3}2} }{n a}
+\frac{1}2(1-3b^2)(1+\Omega_k)
\end{equation}
with $\Omega_{ND}=\frac{\rho_{ND}}{\frac{3H^2}{8 \pi G}}$ as the
density parameter.

\section{\normalsize\bf{Discussion and concluding remarks:}}
In this paper, validity of the second law of thermodynamics on the
event horizon has been analysed for interacting two fluid system.
Here dust is one component of the matter  while the other
component of the matter is holographic dark energy in section II
and in section III the new age graphic dark energy is chosen as
the other component.\\
As there is no explicit expression for the radius of the event
horizon for the new age graphic dark energy model so validity of
the GSLT restricts both the matter as well as the geometry. On the
other hand, for the holographic dark energy model, there exists an
explicit expression for the radius of the event horizon and
consequently, validity of GSLT demands restrictions on matter
only.The restrictions in compact form can be written as the
following:\\

a) \underline{Holographic DE:}\\

$$\omega_D>~max{[-(1+u),u-(1+b^2)(1+u)]}$$
$$or$$
$$\omega_D<~min{[-(1+u),u-(1+b^2)(1+u)]}$$
or equivalently, the parameters $b^2$ and $c$ are constrained as
$$b^2\leq~\frac{\left[u+\frac{2}3\left(1-\frac{\sqrt{\Omega_D-\Omega_k}}c\right)\right]}{(1+u)}~~~and~~~c<\sqrt{\Omega_D-\Omega_k}$$
$$or$$
$$b^2\geq~\frac{\left[u+\frac{2}3\left(1-\frac{\sqrt{\Omega_D-\Omega_k}}c\right)\right]}{(1+u)}~~~and~~~c>\sqrt{\Omega_D-\Omega_k}$$\\

b) \underline{New age graphic DE:}\\

For open or flat model $R_E\geq~R_H=\frac{1}H$ so we have
$$\omega_{ND}>-(1+u)~~i.e.~~b^2\leq
\frac{\left[u+\frac{2\sqrt{\Omega_{ND}}}{3na}\right]}{(1+u)}.$$
However for closed model if $R_E~\geq~R_H$ then the above
inequalities hold, but if $R_E~\leq~R_H$ then the inequalities
will be reversed i.e.
$$\omega_{ND}<-(1+u)~~i.e.~~b^2\geq
\frac{\left[u+\frac{2\sqrt{\Omega_{ND}}}{3na}\right]}{(1+u)}.$$\\
One may note that the second alternative for holographic DE and
the closed model in NAGDE (with $R_E~\leq~R_H$) corresponds to a
possible phantom area.\\
We shall now discuss the validity of GSLT from the present
observational point of view. From the recent observations we have
[25,33,34]
$$\Omega_D=0.72,~~\Omega_k=0.02, a=1~~(present~~ time)~~n=2.7~$$\\

\underline{a)HDE:} The explicit form of $q$ and $\omega_D$ are
given by
$$q=-0.57-1.53b^2$$
$$\omega_D=-1.00005-1.4167b^2$$

Now the validity of GSLT demands $\rho_t+p_t~\geq0$ and
$1+\omega_D^{eff}\geq0$ which gives
$$b^2\leq0.294~~~and~~~c\geq0.8366$$ (which agree with [34])and consequently
$$q~\geq~-1.01982~~~and~~~~\omega_D~\geq~-1.4166.$$\\
Note that the above restriction on $'c'$ agrees with that in
ref.[34]. Thus the holographic dark energy model may be of phantom
 nature for the validity of the GSLT as shown in Fig(I).\\

\begin{figure}

\includegraphics[height=1.5in]{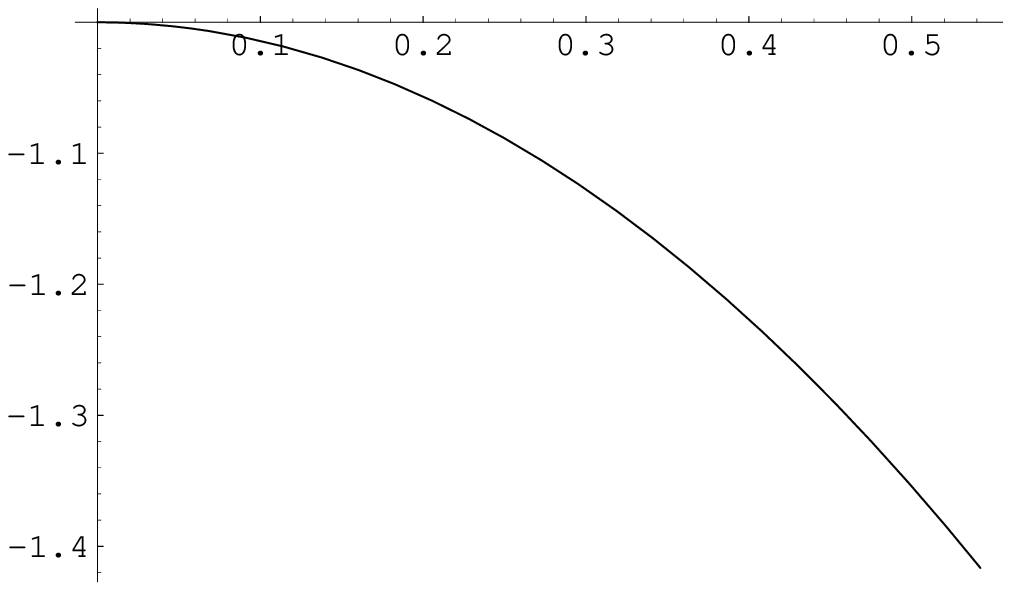}~~~
\includegraphics[height=1.5in]{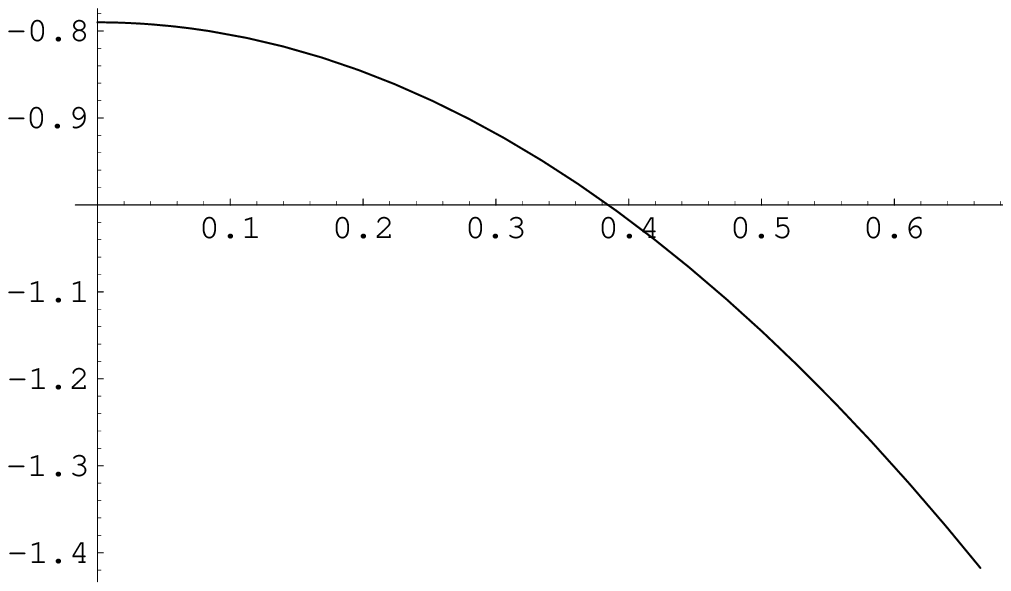}~~~

\vspace{1mm}
Fig.I~~~~~~~~~~~~~~~~~~~~~~~~~~~~~~~~~~~~~~~~~~~~~~~~Fig.II\\

\vspace{5mm} FigI. shows the variation of $\omega_D$ for
holographic dark energy model against the parameter $b$ which
ranges in [0,0.542218] i.e. $b^2\leq0.294$ and FigII. shows the
variation of $\omega_{ND}$ for new age graphic dark energy model
against the parameter $b$ whose range is [0,0.66483] i.e. $b^2
\leq 0.442$ \vspace{6mm}
\end{figure}

\underline{b) NAGDE:} From the observed data, $q$ and
$\omega_{ND}$ are the following $$q=-0.34-1.53b^2$$
$$\omega_{ND}=-0.79-1.42b^2$$
Now  $\rho_t+p_t~\geq0$ gives an upper bound for $'b'$ as $b^2
\leq 0.442$ which gives the explicit restrictions on $q$ and
$\omega_{ND}$ in the following manner:$$q
\geq-1.01626~~~and~~~\omega_{ND}\geq-1.41764.$$\\
Hence in this case also [we can find from Fig(II)] the GSLT may be
valid in quintessence era as well as in phantom era. Therefore,
the phantom divide line has no influence on the validity of the
GSLT in both the dark energy models i.e. there may be a smooth
transition between the quintessence and phantom era in the new age
graphic dark energy model.

Finally, we note that to examine the validity of GSLT we have not
used any explicit form of entropy or temperature at the event
horizon only we have imposed the condition that the  first law of
thermodynamics is valid here which may be considered as an
conservation equation ( since we have assumed that the universe is
in thermal equilibrium). However in Ref.[7] it has been shown that
the usual definition of the temperature (Hawking temperature) and
entropy (Bekenstein entropy) do not hold on the event horizon and
consequently the first law of thermodynamics is not satisfied on
the event horizon. Therefore, for future work we examine the
validity of the first law thermodynamics with appropriate choice
of entropy and temperature on the event horizon.\\

{\bf References:}\\
\\
$[1]$ S.W.Hawking, \textit{Commun.Math.Phys} \textbf{43} 199
(1975).\\\\
$[2]$ J.D.Bekenstein, \it{Phys. Rev. D} {\bf 7} 2333 (1973).\\\\
$[3]$ J.M.Bardeen , B.Carter and S.W.Hawking, {\it
Commun.Math.Phys } {\bf 31} 161 (1973).\\\\
$[4]$ T.Jacobson, \it {Phys. Rev Lett.} {\bf 75} 1260 (1995). \\\\
$[5]$ T.Padmanabhan, \it {Class. Quantum Grav} {\bf 19} 5387
(2002); \it{Phys.Rept} {\bf 406} 49 (2005).\\\\
$[6]$ R.G. Cai and L.M. Cao , {\it Phys. Rev. D} {\bf 75} 064008 (2007).\\\\
$[7]$ B. Wang, Y. Gong, E. Abdalla , \it{Phys. Rev. D} {\bf 74}
083520 (2006).\\\\
$[8]$ S.Perimutter et. al. [Supernova Cosmology Project
Collaboration] \it{Astrophys. J} {\bf 517} (1999) 565; A.G. Riess
et. al. [Supernova Search Team Collaboration] \it{Astron. J} {\bf
116} (1998) 1009; P. Astier et. al. [The SNLS Collaboration]
\it{Astron. Astrophys.} {\bf 447} (2006) 31; A.G. Riess et. al.
\it{Astrophys. J} {\bf 659} (2007) 98.\\\\
$[9]$ D.N. Spergel et. al. [WMAP Collaboration] \it{Astrophys. J
Suppl.} {\bf 148} (2003) 175; H.V. Peiris et. al. [WMAP
Collaboration] \it{Astrophys. J. Suppl.} {\bf 148} (2003) 213;
D.N. Spergel et. al. [WMAP Collaboration] \it{Astrophys. J Suppl.}
{\bf 170} (2007) 377; E. Komatsu et. al. [WMAP Collaboration]
arXiv:0803.0547 [astro-ph].\\\\
$[10]$ V. Sahi , {\it AIP Conf. Proc.} {\bf 782} (2005) 166 ; [J. Phys. Conf. Ser. {\bf 31} (2006) 115].\\\\
$[11]$ T. Padmanavan , {\it Phys. Rept. } {\bf 380} (2002) 235.\\\\
$[12]$ E.J. Copeland, M. Sami and S. Tsujikawa , {\it IJMPD } {\bf 15} (2006) 1753.\\\\
$[13]$ R. Durrer and R. Marteens , {\it Gen. Rel. Grav.} {\bf 40}
(2008) 301.\\\\
$[14]$ S. Nojiri and S.D. Odintsov , {\it Int. J. Geom. Meth. Mod.
Phys.} {\bf 4} (2007) 115.\\\\
$[15]$ S. Nojiri , S. Odintsov , {\it arXiv: 0801.4843} {\bf
[astro-ph]};  S. Nojiri , S. Odintsov , {\it arXiv: 0807.0685}
{\bf [hep-th ]}; S. Capozziello, {\it IJMPD} {\bf 11} (2002) 483.\\\\
$[16]$ S.M. Carroll , V. Duvvuri, M. Trodden and M.S. Turner {\it
Phys. Rev. D} {\bf 68} (2004) 043528.\\\\
$[17]$ S. Nojiri and S.D. Odintsov , {\it Phys. Rev. D} {\bf 68}
(2003) 123512; S. Nojiri and S.D. Odintsov , {\it Phys. Rev. D} {\bf 74} (2006) 086005.\\\\
$[18]$ Kazuharu Bamba and Chao-Qiang Geng , {\it arXiv: 0901.1509}
{\bf [hep-th]}.\\\\
$[19]$ H.M. Sadjadi , {\it Phys. Rev. D} {\bf 76} (2007)
104024.\\\\
$[20]$ A.G. Cohen , D.B. Kaplan and A.E. Nelson , {\it Phys. Rev.
Lett.} {\bf 82} 4971 (1999);\\\\
$[21]$ Q.G. Huang and M. Li , {\it JCAP} {\bf 0408} 013 (2004);\\\\
$[22]$ M. Li , {\it Phys. Lett. B} {\bf 603} 01 (2004);\\\\
$[23]$ H.Wei, R.G.Cai, {\it Phys. Lett. B.} {\bf 660} 113
(2008);\\\\
$[24]$ R.G.Cai, {\it Phys. Lett. B.} {\bf 657} 228 (2007);\\\\
$[25]$ H.Wei, R.G.Cai, {\it Phys. Lett. B.} {\bf 663} 1
(2008);\\\\
$[26]$ A.G. Cohen , D.B. Kaplan and A.E. Nelson , {\it Phys. Rev.
Lett.} {\bf 82} 4971 (1999); P.Horava, D.Minic, {\it Phys. Rev.
Lett.} {\bf 85} 1610 (2000); S.D.Thomas ,{\it Phys. Rev. Lett.}
{\bf 89} 081 301 (2002);M. Li , {\it Phys. Lett. B} {\bf 603} 01 (2004);\\\\
$[27]$ M.R. Setare , {\it JCAP} {\bf 01} 023 (2007);\\\\
$[28]$ N. Mazumder and S. Chakraborty , {\it Gen.Rel.Grav.}{\bf 42} 813 (2010);\\\\
$[29]$ R. G. Cai and S. P. Kim, {\it JHEP} {\bf 02} 050 (2005).\\\\
$[30]$ R.S. Bousso, \it{Phys. Rev. D} {\bf 71} 064024 (2005).\\\\
$[31]$ G. Izquierdo and D. Pavon, {\it Phys. Lett. B} {\bf 633}
420 (2006).\\\\
$[32]$ N. Mazumder and S. Chakraborty, {\it Class. Quant. Gravity}
{\bf 26} 195016 (2009).\\\\
$[33]$ C.L. Bennett et al., {\it Astrophys. J. Suppl.} {\bf 148},
1 (2003); D.N. Spergel, {\it Astrophys. J. Suppl.} {\bf 148}, 175
(2003); M. Tegmark et al., {\it Phys. Rev. D} {\bf 69}, 103501
(2004); U. Seljak, A. Slosar, P. McDonald, {\it J. Cosmol.
Astropart. Phys.} {\bf 10}, 014 (2006); D.N. Spergel et al., {\it
Astrophys. J. Suppl.} {\bf 170}, 377 (2007).\\\\
$[34]$ X. Zhang and F. Q. Wu, {\it Phys. Rev. D} {\bf 72}, 043524
(2005) [astro-ph/0506310].\\\\

\end{document}